\documentclass[dvips]{acta}
\usepackage{supertabular,lscape,epsfig}
\usepackage{amssymb}
\usepackage{amsmath}

\SetPages{0}{0}

\SetVol{0}{2014}

\begin{document}

\begin{Titlepage}
\Title{Revisiting parameters for the WASP-1 planetary system\footnote{Partly based on data collected with telescopes at the Rozhen National Astronomical Observatory, and observations obtained with telescopes of the University Observatory Jena, which is operated by the Astrophysical Institute of the Friedrich-Schiller-University.}}
\Author{G.~~M~a~c~i~e~j~e~w~s~k~i$^{1}$, ~ J.~~O~h~l~e~r~t$^{2,3}$, ~ D.~~D~i~m~i~t~r~o~v$^{4}$, ~ D.~~P~u~c~h~a~l~s~k~i$^{1}$, ~ J.~~Nedoro\v{s}\v{c}\'{i}k$^{5}$, ~M.~~Va\v{n}ko$^{5}$, ~ C.~~M~a~r~k~a$^{6}$,  ~ S.~~B~a~a~r$^{6}$, ~ St.~~R~a~e~t~z$^{6,7}$, ~ M.~~S~e~e~l~i~g~e~r$^{6}$,  ~ and ~ R.~~N~e~u~h~\"a~u~s~e~r$^{6}$}
{$^1$Centre for Astronomy, Faculty of Physics, Astronomy and Informatics, 
         Nicolaus Copernicus University, Grudziadzka 5, 87-100 Torun, Poland\\ 
         e-mail: gm@astri.umk.pl\\
 $^2$Michael Adrian Observatorium, Astronomie Stiftung Trebur, 65468 Trebur, Germany\\
 $^3$University of Applied Sciences, Technische Hochschule Mittelhessen, 61169 Friedberg, Germany\\
 $^4$Institute of Astronomy, Bulgarian Academy of Sciences, 72 Tsarigradsko Chausse Blvd., 1784 Sofia, Bulgaria\\
 $^5$Astronomical Institute of the Slovak Academy of Sciences, 059 60 Tatransk\'a Lomnica, Slovakia\\
 $^6$Astrophysikalisches Institut und Universit\"ats-Sternwarte, 
         Schillerg\"asschen 2--3, 07745 Jena, Germany\\ 
 $^7$European Space Agency, ESTEC, SRE-S, Keplerlaan 1, 2201 AZ Noordwijk, The Netherlands} 

\Received{January 30, 2014}
\end{Titlepage}

\Abstract{We present thirteen new transit light curves for the WASP-1~b exoplanet. Observations were acquired with 0.5 - 1.2-m telescopes between 2007 and 2013. Our homogeneous analysis, which also includes the literature data, results in determining precise system parameters. New values are in agreement with those reported in previous studies. Transit times follow a linear ephemeris with no sign of any transit time variations. This finding is in line with the paradigm that Jupiter-like planets on tight orbits are devoid of close planetary companions.}{planetary systems -- stars: individual: WASP-1 -- planets and satellites: individual: WASP-1~b}


\section{Introduction}

Observations of transiting exoplanets offer a wealth of knowledge about extrasolar planetary systems. Photometric and spectroscopic datasets, when combined together, allow the planetary radius and exact planetary mass to be determined. These quantities give the mean planetary density, which is a clue to the internal structure of a planet (\eg Nettelmann \etal 2010). Observations of occultations of transiting planets (\eg Snellen \etal 2009) constrain an orbital eccentricity, which is usually poorly determined from radial velocity (RV) measurements alone. These observations also allow one to determine chemical properties of the planetary atmosphere with the spectrophotometry technique (\eg Charbonneau \etal 2008). Observations of the Rossiter--McLaughlin effect provide the sky-projected angle between the stellar spin axis and the planetary orbital axis (\eg Queloz \etal 2000). This information gives important insight into theories of planet formation and dynamical evolution. 

The architecture of planetary systems with a transiting planet may be studied with the RV and transit timing methods to search for additional planets. The former technique has already brought discoveries of companions on wide orbits (\eg Bakos et al. 2009). The latter method, based on the detection of transit time variations (TTVs,  \eg Agol \etal 2005, Holman \& Murray 2005), is sensitive to terrestrial-mass planetary companions close to low-order mean-motion resonances (MMRs, Steffen \etal 2007). In such configurations, the mutual gravitational interaction of planets may generate deviations from strictly periodic transit times. The terrestrial-mass planets are usually below the current RV sensitivity limit which makes the TTV method, {\bf in addition to gravitational microlensing}, the unique discovery tool. So far, the TTV technique has been successfully applied to data from the space-based {\it Kepler} transit survey (Borucki \etal 2010). In transiting multi-planetary systems, analyses of TTVs have allowed one to confirm or validate the planetary nature of many candidates (\eg Holman \etal 2010). In single transiting planet systems, dynamical analyses of observed transit times have led to discoveries of unseen planetary companions (\eg Ballard \etal 2011), some of them confirmed independently by RV measurements (Barros \etal 2014).

Statistical studies of hot-Jupiter candidates (planets with masses of the order of 1 Jupiter mass, $M_{\rm{Jup}}$, and orbital periods shorter than 5 days) in the {\it Kepler} sample show that these planets accompanied by nearby low-mass planets are rare (Steffen \etal 2012, Steffen \& Farr 2013). In contrast, planets on slightly wider orbits, so called warm Jupiters, Neptunes, or super Earths, are observed to form multi-planetary systems with planetary pairs close to MMRs (\eg Ming \etal 2013). The relative isolation of hot Jupiters could be the result of the initial formation conditions and dynamical evolution of their planetary systems (Kley \& Nelson 2012). Theoretical studies show that the system architecture may depend on many factors, including the mass ratio of planets relative to the central star, the mass of the protoplanetary disk, and the mechanism of the planetary migration (Goldreich \& Schlichting 2013). Thus, the detection or a lack of companion planets close to hot Jupiters may put forward observational constraints on the theory of planet formation (\eg Ogihara \etal 2013; see also Steffen \etal 2012 for a detailed discussion). 

It has been shown that a terrestrial-mass planet perturbing a hot-Jupiter gas giant in an orbital resonance can produce a TTV signal of an amplitude of a few minutes (Steffen \etal 2007). Such a signal can be detected using timing observations acquired with 0.5--1.0-m telescopes. Being encouraged by this finding, we have organized observing campaigns for selected transiting planets to search for their possible TTVs (Maciejewski \etal 2011a). So far, we have put constraints on the upper mass of additional planets in WASP-3 and TrES-3 systems (Maciejewski \etal 2013b, Va\v{n}ko \etal 2013), refined transit ephemerides for some transiting planets (Maciejewski \etal 2011c, 2013c, Raetz \etal 2014), studied physical properties of the WASP-10~b planet and the activity of its host star (Maciejewski \etal 2011b), and finally found a hint of orbital perturbations for the WASP-12~b planet (Maciejewski \etal 2013a). In this paper, we study the transiting exoplanet WASP-1~b (Collier Cameron \etal 2007). Our observing campaign, spanning six years, resulted in refining system parameters and putting upper-mass constraints on the second planet in the system.

The WASP-1 star (GSC 02265-00107, $V=11.31$ mag) with its transiting planet belongs to the first planetary systems discovered by the SuperWASP survey (Pollacco \etal 2006). The planetary nature of its companion was confirmed with RV measurements acquired with the SOPHIE spectrograph (Bouchy \etal 2006). The planet was found to be bloated with a sub-Jovian mass and the radius 1.4 times greater than that of Jupiter. It orbits a dwarf star of the F7 spectral type on a 2.5-d orbit with the semi-major axis of 0.038~AU. The system parameters were refined by follow-up observations by Charbonneau \etal (2007) and Shporer \etal (2008) and a re-analysis of Southworth (2008; 2010). Szab\'o \etal (2010) refined the transit ephemeris with new transit observations acquired with a small telescope. 

The host star was the subject of a detailed spectroscopic study by Stempels \etal (2007). Precise RV measurements, obtained throughout a transit by Simpson \etal (2011) and Albrecht \etal (2011), show that WASP-1~b belongs to the group of misaligned planetary systems with the almost polar orbit. Wheatley \etal (2010) observed an occultation of WASP-1~b with the Spitzer Space Telescope. Their results show that there is a strong temperature inversion in the planet's atmosphere. Furthermore, the phase of mid-occultation was found to be consistent with a value of 0.5. This finding suggests that the planetary orbit is circularised.

There is a number of mechanisms proposed to explain anomalous sizes of inflated exoplanets (see \eg Martin \etal 2011 for references). Early studies suggested that a small but non-zero value of the orbital eccentricity of WASP-1~b, excited by an undetected planetary companion, could bloat the planet through tidal heating (Mardling 2007). This finding motivated us to study the WASP-1 system through the transit timing of WASP-1~b.


\section{Observations and data reduction}

New transit light curves were acquired using six instruments with diameters of the main mirror between 0.5 and 1.2 m, located in Germany, Bulgaria, Poland, Slovakia, and Namibia (see Table 1 for details). All instruments were equipped with charge coupled devices (CCDs), thus standard data reduction procedure, including debiasing, dark correction, and flat-fielding using sky flats, was applied. Differential aperture photometry was performed with respect to the comparison stars available in a field of view. The size of aperture was optimized to achieve the smallest scatter in the out-of-transit light curves. To account for photometric trends, mainly caused by differential atmospheric extinction and different spectral types of the target and comparison stars, the light curves were de-trended by fitting a second-order polynomial function of time along with a trial transit model. In the first iteration, complete light curves were de-trended
using the trial transit model, the initial parameters of which were taken from the literature. A preliminary transit model, based on complete light curves, was obtained following a procedure described in Section 3.1. In the second iteration, partial transit light curves were de-trended employing a second-order polynomial function and the preliminary transit model. The de-trending procedure was performed with the {\sc jktebop} code (Southworth \etal 2004a,b), which allows photometric trends to be modeled as polynomials up to 5th order. The best-fitting trend was subtracted from each light curve. 

\MakeTable{c l l  c}{12.5cm}{List of telescopes, which participated in the observing campaign, sorted according to the mirror diameter. CCD size is the number of pixels and pixel size. FoV is the field of view of the instrument. $N_{\rm{tr}}$ is the number of useful light curves.}
{\hline
\# & Telescope    &  Detector  &  $N_{\rm{tr}}$\\
     & Observatory &  CCD size &    \\
     &  Location     &  FoV         &    \\
\hline
1 & 1.2-m Cassegrain                & Roper Scientific EEV 1340EMB$^{*}$  & 1 \\
  & Michael Adrian Observatory  & $1340 \times 1300$, 20 $\mu$m &   \\
  & Trebur, Germany                  & $9'.7 \times 9'.4$ & \\

  &                  &  SBIG STL-6303E$^{**}$ & 2 \\
  &    & $1536\times1024$, 18.0 $\mu$m  &   \\
  &                    & $10'.0 \times 6'.7$ & \\

2 & 0.9/0.6-m Schmidt$^{***}$      & CCD-imager STK & 0 \\
  & University Observatory Jena  & $2048\times 2048$, 13.5 $\mu$m &  \\
  &  Gro{\ss}schwabhausen near Jena, Germany & $52'.8 \times 52'.8$   &  \\

3 & 0.6-m Cassegrain                           &  FLI PL09000 & 4 \\
  &  National Astronomical Observatory & $3056 \times 3056$, 12 $\mu$m & \\
  & Rozhen, Bulgaria                             & $17'.3 \times 17'.3$   &   \\

4 & 0.6-m Cassegrain                                             &  SBIG STL-1001 & 4 \\
  & Toru\'n Centre for Astronomy  & $1024 \times 1024$, 24 $\mu$m &  \\
  & Piwnice near Toru\'n, Poland & $11'.8\times11'.8$  &  \\

5 & 0.6-m Cassegrain                           &  Moravian Instruments G4-9000 & 1 \\
  &  Star\'a Lesn\'a Observatory & $3048 \times 3048$, 12 $\mu$m & \\
  & Star\'a Lesn\'a, Slovakia                             & $16' \times 16'$   &   \\

6 & 0.5-m Cassegrain                           &  SBIG ST-10 & 1 \\
  &  International Amateur Observatory & $2184 \times 1472$, 6.8$ \mu$m & \\
  & Hakos, Namibia                             & $11'.3 \times 7'.6$   &   \\
\hline
\multicolumn{4}{l}{$^{*}$ up to 2008 }\\
\multicolumn{4}{l}{$^{**}$ from 2010 }\\
\multicolumn{4}{l}{$^{***}$~see Mugrauer \& Berthold (2010) for details }\\
}

Magnitudes were transformed into fluxes and normalised to have a mean of unity outside of the transit. The timestamps in geocentric Julian dates in coordinated universal time (UTC) were converted to barycentric Julian dates in barycentric dynamical time (Eastman \etal 2010). {\bf To ensure a reliable time survey and exclude systematic errors in the recorded times, local computer clocks were synchronized with the network time protocol software, usually accurate to better than 0.1\,s.} To quantify the quality of each light curve, we used the photometric noise rate ($pnr$, Fulton \etal 2011) defined as 
\begin{equation}
  pnr = \frac{rms}{\sqrt{\Gamma}}
\end{equation}
where the root mean square of the residuals, $rms$, is calculated from the light curve and a fitted model, and $\Gamma$ is the median number of exposures per minute.

We acquired 13 new transit light curves\footnote{The data are available in a machine-readable form at http://ttv.astri.umk.pl.} for WASP-1~b between 2007 September and 2013 November, summarised in Table~2. Some light curves are incomplete because they were affected by adverse weather conditions or limited visibility of the target above the horizon. The observations were mostly collected in $R$ and $V$ filters. Some data were acquired without any filter (in so called ''clear'' filter or white light) with a spectral band determined by instrumental response, stellar spectral energy distribution, and the wavelength-depended sky transmission. This method is expected to increase instrument efficiency and to reduce photometric noise, hence to provide more precise mid-transit times. One must note, however, that a poorly determined efficient band, which in addition may vary due to variable atmospheric conditions, introduces uncertainties to the coefficients of the limb darkening (LD) law. In consequence, white-light data should be treated with caution while determining system parameters. In this study, we used those data for transit timing only.

\MakeTable{r l l l c c}{12.5cm}{New transit light curves acquired for WASP-1~b: date UT is given for the middle of the transit, epoch is the transit number from the initial ephemeris given in Collier Cameron \etal (2007), $X$ is the airmass change during transit observations, $\Gamma$ is the median number of exposures per minute, $pnr$ is the photometric scatter in millimagnitudes per minute of observation.}
{\hline
\# & Date UT (epoch) &  Telescope & Filter  & $\Gamma$ & $pnr$\\ 
    &     &    $X$       & Sky conditions &  &\\ 
\hline
   1 & 2007 Sep 10 (175) & Hakos 0.5 m  & $R_{\rm{J}}$ & 1.36  &  3.39\\ 
      &               &  $1.86\rightarrow2.78$  & Clear &  &  \\ 
   2 & 2008 Sep 27 (327) & Trebur 1.2 m & $R_{\rm{C}}$ & 1.33 & 1.53 \\
      &  & $1.11\rightarrow1.05\rightarrow1.63$ & Mostly clear & &  \\
   3 & 2008 Dec 22 (361) & Piwnice 0.6 m & $R_{\rm{C}}$ & 4.29 & 2.78 \\ 
      & & $1.08\rightarrow1.23$ & Clear & &  \\ 
   4 & 2009 Aug 02 (450) & Rozhen 0.6 m & $V$ & 0.83 & 2.47\\ 
      & & $1.71\rightarrow1.01\rightarrow1.02$ & Clear & &  \\ 
   5 & 2009 Sep 29 (473) & Rozhen 0.6 m & $R_{\rm{C}}$ & 0.83 & 2.91\\ 
      & & $1.06\rightarrow1.01\rightarrow1.36$ & Clear & &  \\ 
   6 & 2009 Nov 21 (494) & Rozhen 0.6 m & $R_{\rm{C}}$ & 0.95 & 2.52\\ 
      &  & $1.02\rightarrow1.01\rightarrow2.11$ & Clear & &  \\ 
   7 & 2010 Oct 17 (625) & Trebur 1.2 m & $R_{\rm{B}}$ & 2.14 & 1.50\\
      & & $1.07\rightarrow1.05\rightarrow1.72$ & Clear & &  \\
  8 & 2013 Sep 07 (1044) & Piwnice 0.6 m & ''clear'' & 0.86 & 3.04 \\ 
     &  &  $1.74\rightarrow1.09$ & Occasional high clouds & &  \\ 
   9 & 2013 Oct 03 (1054) & Trebur 1.2 m & ''clear'' & 1.82 & 1.20\\
      & & $1.08\rightarrow1.05\rightarrow1.82$ & Mostly clear & &  \\
 10 & 2013 Oct 03 (1054) & Piwnice 0.6 m & ''clear'' & 2.00 & 2.41 \\ 
      & &  $1.07\rightarrow1.92$ & Clear, high humidity & &  \\ 
 11 & 2013 Oct 08 (1056) & Star\'a Lesn\'a 0.6 m & $R_{\rm{C}}$ & 0.89 & 4.10\\ 
      & & $1.05\rightarrow1.56$ & Clear & &  \\ 
  12 & 2013 Nov 09 (1069) & Rozhen 0.6 m & $R_{\rm{C}}$ & 0.65 & 2.72 \\ 
       &  & $1.09\rightarrow1.01\rightarrow1.31$ & Clear & &  \\ 
 13 & 2013 Nov 09 (1069) & Piwnice 0.6 m & ''clear'' & 2.61 & 1.63\\ 
      &  & $1.18\rightarrow1.07\rightarrow1.12$ & High clouds & &  \\ 
\hline
}


\section{Results}


\subsection{Transit model}\label{sec:Transits}

Transit light curves were modeled with the Transit Analysis Package\footnote{http://ifa.hawaii.edu/users/zgazak/IfA/TAP.html} ({\sc TAP}, Gazak \etal 2012). The best-fitting parameters of a transit, which is based on the model of  Mandel \& Agol (2002), are found with the  Markov Chain Monte Carlo (MCMC) method, employing the Metropolis-Hastings algorithm and a Gibbs sampler. The conservative uncertainty estimates are derived with the wavelet-based technique of Carter \& Winn (2009) that accounts for the time-correlated (red) noise in the modeled data.  

A set of 19 light curves, including a light curve from Charbonneau \etal (2007), two light curves from Shporer \etal (2008), two partial light curves from Albrecht \etal (2011), a high quality light curve provided by F.Harmuth to the Exoplanet Transit Database (ETD, Poddan\'y \etal 2010), and 13 new light curves presented in this paper, were modeled simultaneously to improve the determination of system parameters. The orbital inclination $i_{\rm{b}}$, the semimajor-axis scaled by stellar radius $a_{\rm{b}}/R_{*}$, and the planetary to stellar radii ratio $R_{\rm{b}}/R_{*}$ were linked together for all light curves in Sloan-$z$, $V$, $R$, and $I$ filters. For light curves acquired in ''clear'' filter, their morphology may be affected by systematics caused by not precisely known LD coefficients, so their parameters were allowed to vary independently. The only restriction was put on $i_{\rm{b}}$, which was allowed to vary around the best fitting value under Gaussian penalty with the 1-$\sigma$ uncertainty of the parameter. The orbital period was fixed to a value obtained in Sect.~3.2. Mid-transit times of individual light curves were free parameters to account for possible timing variations. Two transits were observed simultaneously with two different telescopes. In those cases, the mid-transit times were linked together to obtain a single determination for a given epoch. {\bf It is worth noticing that individual mid-transits times for both epochs were found to be consistent well within $1\sigma$ when they were allowed to be determined independently in a test run. This observation allowed us to exclude any systematic errors in the time survey for both epochs.}

To parametrise the flux distribution across the stellar disk, TAP employs a quadratic LD law in a form
\begin{equation}
     I_{\theta}/I_0 = 1 - u_1 (1 - \cos\theta) - u_2 (1 - \cos\theta)^2\, , \; 
\end{equation}
where  $I_{\theta}$ is the intensity at angle $\theta$ which is the angle between the surface normal and the line of sight, $I_0$ is the intensity at the centre of the stellar disk, and $u_1$ and $u_2$ are linear and quadratic LD coefficients, respectively. For individual filters, the values of $u_1$ and $u_2$ coefficients were linearly interpolated from tables of Claret \& Bloemen (2011) with an on-line tool\footnote{http://astroutils.astronomy.ohio-state.edu/exofast/limbdark.shtml} of the \textsc{EXOFAST} applet  (Eastman \etal 2013). The efficient maximum of energy distribution in the ''clear'' filter was found to be between $V$ and $R$ bands. Thus, the LD coefficients for white-light data were calculated as average values of those in $V$ and $R$ bands. We notice that this approach results in values close to bolometric LD coefficients. The parameters of the host star (effective temperature, surface gravity, and metallicity) were taken from Torres \etal (2012). The photometric dataset was found to be not precise enough to allow LD coefficients to be free parameters of the model. To take uncertainties of the theoretical LD law into account, LD coefficients were allowed to vary under the Gaussian penalty of $\sigma = 0.05$. 

A circular orbit of WASP-1~b was assumed as discussed in Sect.~3.3. The fitting procedure was also allowed to account for possible linear trends in individual light curves and to include their uncertainties in a total error budget of the fit. 

In the final iteration, ten MCMC chains, each containing $10^6$ steps, were computed. The first 10\% of the results were discarded from each chain to minimise the influence of the initial values of the parameters. The best-fitting parameters, determined as the median values of marginalised posteriori probability distributions, are given in Table~3. Their upper and lower 1\,$\sigma$ errors are determined by 15.9 and 84.1 percentile values of the distributions. The transit parameter $b_{\rm{b}}$, defined as
\begin{equation}
     b_{\rm{b}}=\frac{a_{\rm{b}}}{R_{*}}\cos{i_{\rm{b}}}\, , \; 
\end{equation}
and the mean stellar density $\rho_{*}$, which can be directly calculated from other parameters with a formula
\begin{equation}
     \rho_{*} = \left(\frac{2\pi}{P_{\rm{b}}}\right)^2 \left(\frac{a_{\rm{b}}}{R_{*}}\right)^3\, , \; 
\end{equation}
are also given. For comparison, we also include determinations from previous studies. The individual light curves with models and residuals are plotted in Fig.~1. 

\begin{figure}[thb]
\begin{center}
\includegraphics[width=1.0\textwidth]{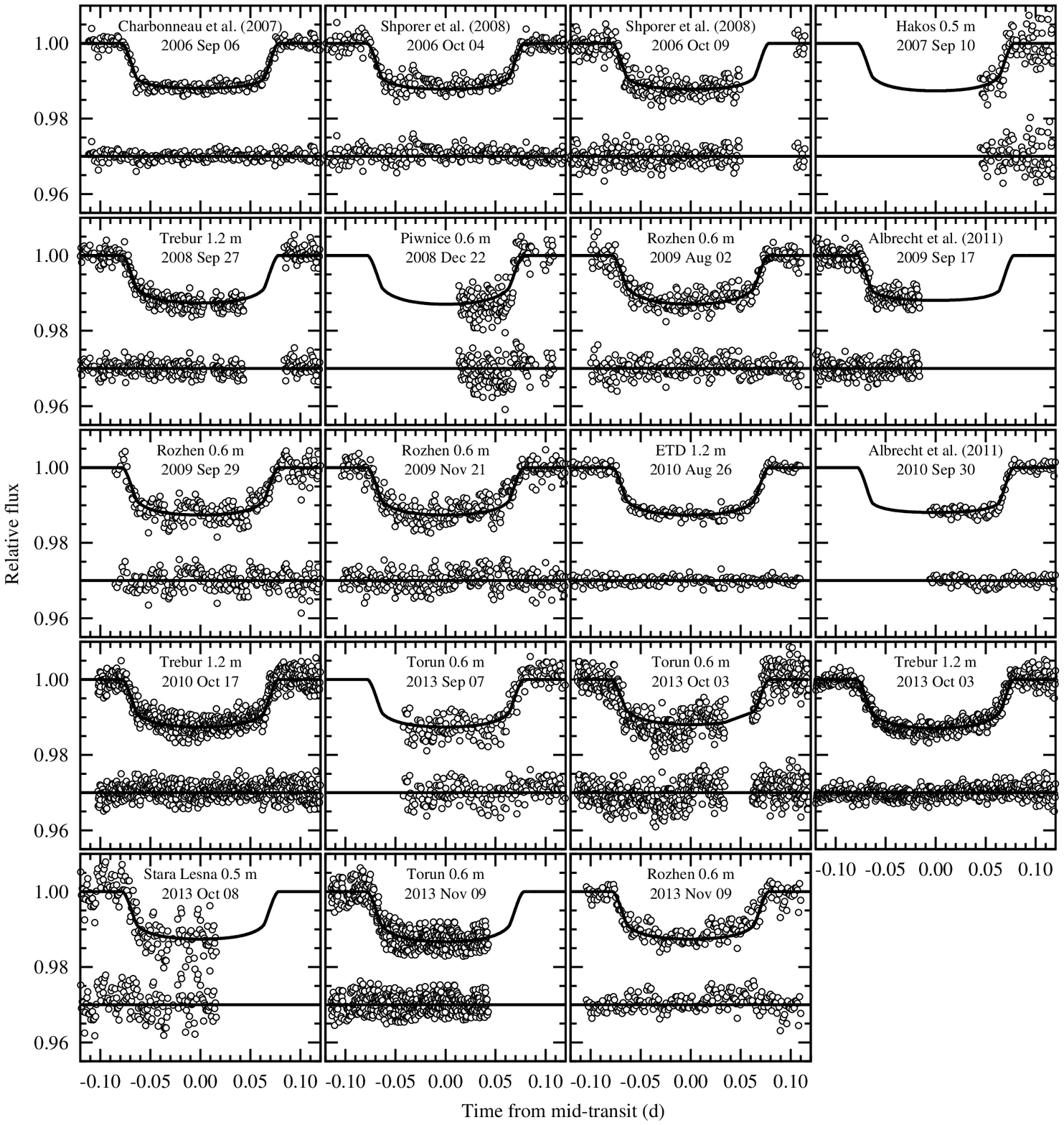}
\end{center}
\FigCap{New and published transit light curves for WASP-1~b, sorted by observation dates. The best-fit model is plotted with continuous lines. The residuals are plotted in bottom parts of the panels.}
\end{figure}

\begin{landscape}
\MakeTable{lccccccc}{18cm}{Parameters of the WASP-1 system.}
{\hline
Parameter &  This work & Col07+Ste07 & Cha07  & Shp08 & Sou08-10 & Sim11 & Alb11  \\
\hline
\multicolumn{8}{l}{Planetary properties}\\
$R_{\rm{b}}/R_{*}$ & $0.10362^{+0.00074}_{-0.00075}$ & 0.093--0.104   & $0.10189\pm0.00093$  & $0.102\pm0.001$  & $0.1033\pm0.0016$ & $0.10271\pm0.00058$  & $0.1059\pm0.0006$ \\
$M_{\rm{b}}$ ($M_{\rm{Jup}}$) & $0.854^{+0.057}_{-0.052}$ & $0.79^{+0.13}_{-0.06}$ & & (0.80--1.05)$_{-0.07}^{+0.09}$ & & & \\
$R_{\rm{b}}$ ($R_{\rm{Jup}}$) & $1.483^{+0.024}_{-0.034}$ & 1.33--2.53 & $1.443\pm0.039$ & (1.36--1.49)$\pm0.06$ & & & \\
$\rho_{\rm{b}}$ ($\rho_{\rm{Jup}}$) & $0.262^{+0.022}_{-0.024}$ & & & & & & \\
$g_{\rm{b}}$ (m~s$^{-2}$) & $9.8^{+0.8}_{-0.9}$ & & & & & \\
\multicolumn{8}{l}{Orbital properties}\\
$i_{\rm{b}}$ ($^{\circ}$) & $90.0\pm1.3$ & & $>86.1$ & & $>84.0$ & $88.65^{+0.51}_{-0.55}$ & $90\pm2$ \\
$b_{\rm{b}}$ & $0.00\pm0.12$ & 0.0--0.8 & $<0.336$ &  $0.03\pm0.17$ &  & & \\
$k_{\rm{b}}$ (m~s$^{-1}$) & $110.9\pm6.1$ & $115\pm11$ & & & & $109.4\pm5.6$ & \\
$a_{\rm{b}}$ (AU) & $0.03889^{+0.00053}_{-0.00073}$ & $0.0379\pm0.0042$ & & & & & \\
\multicolumn{8}{l}{Stellar properties}\\
$a_{\rm{b}}/R_{*}$ & $5.687^{+0.034}_{-0.061}$ & 3.85--5.95 & & 5.53--5.99  & $5.50^{+0.39}_{-0.33}$  & $5.64\pm0.13$  & 5.68--8.81  \\
$M_{*}$ ($M_{\odot}$) & $1.24^{+0.05}_{-0.07}$ & $1.15^{+0.24}_{-0.09}$ & & & & & \\
$L_{*}$ ($L_{\odot}$) & $2.76^{+0.26}_{-0.24}$ & & & & & & \\
$R_{*}$  ($R_{\odot}$) & $1.470^{+0.022}_{-0.032}$ & $1.24^{+0.68}_{-0.20}$ & $1.453\pm0.032$ & (1.38--1.51)$\pm0.06$ & & & \\
$\rho_{*}$  $(\rho_{\odot})$ & $0.389^{+0.007}_{-0.012}$ & & & & & & \\
$\log g_{*}$ (cgs units) & $4.195^{+0.010}_{-0.017}$ & $4.28\pm0.15$ & & & & $4.19\pm0.07$ \\
age (Gyr) & $3.4^{+1.2}_{-0.6}$ & $2.0\pm1.0$ & & & & & \\
\hline
\multicolumn{8}{p{18cm}}{References: Col07 - Collier Cameron \etal (2007), Ste07 - Stempels \etal (2007), Cha07- Charbonneau \etal (2007), Shp08 - Shporer \etal (2008), Sou08-10 - Southworth (2008; 2010), Sim11 - Simpson \etal (2011), Alb11 - Albrecht \etal (2011).}
}
\end{landscape}

An analysis of relations between fitted parameters (Fig.~2) reveals a statistically significant correlation between $i_{\rm{b}}$ and $a_{\rm{b}}/R_{*}$. 
This is a typical behaviour when there are no other constraints on one of these parameters (see \eg Hoyer \etal 2012). However, we notice that keeping $a_{\rm{b}}/R_{*}$ fixed on the value based on spectroscopic determinations results in noticeable underestimated uncertainties of $i_{\rm{b}}$, compared to our final solution. Our final approach is more conservative and provides an independent determination of $a_{\rm{b}}/R_{*}$. Remaining parameters, especially mid-transit times of individual light curves, reveal no dependence between each other. Histograms of MCMC parameters (Fig.~3) show unimodal distributions.  

\begin{figure}[htb]
\begin{center}
\includegraphics[width=0.8\textwidth]{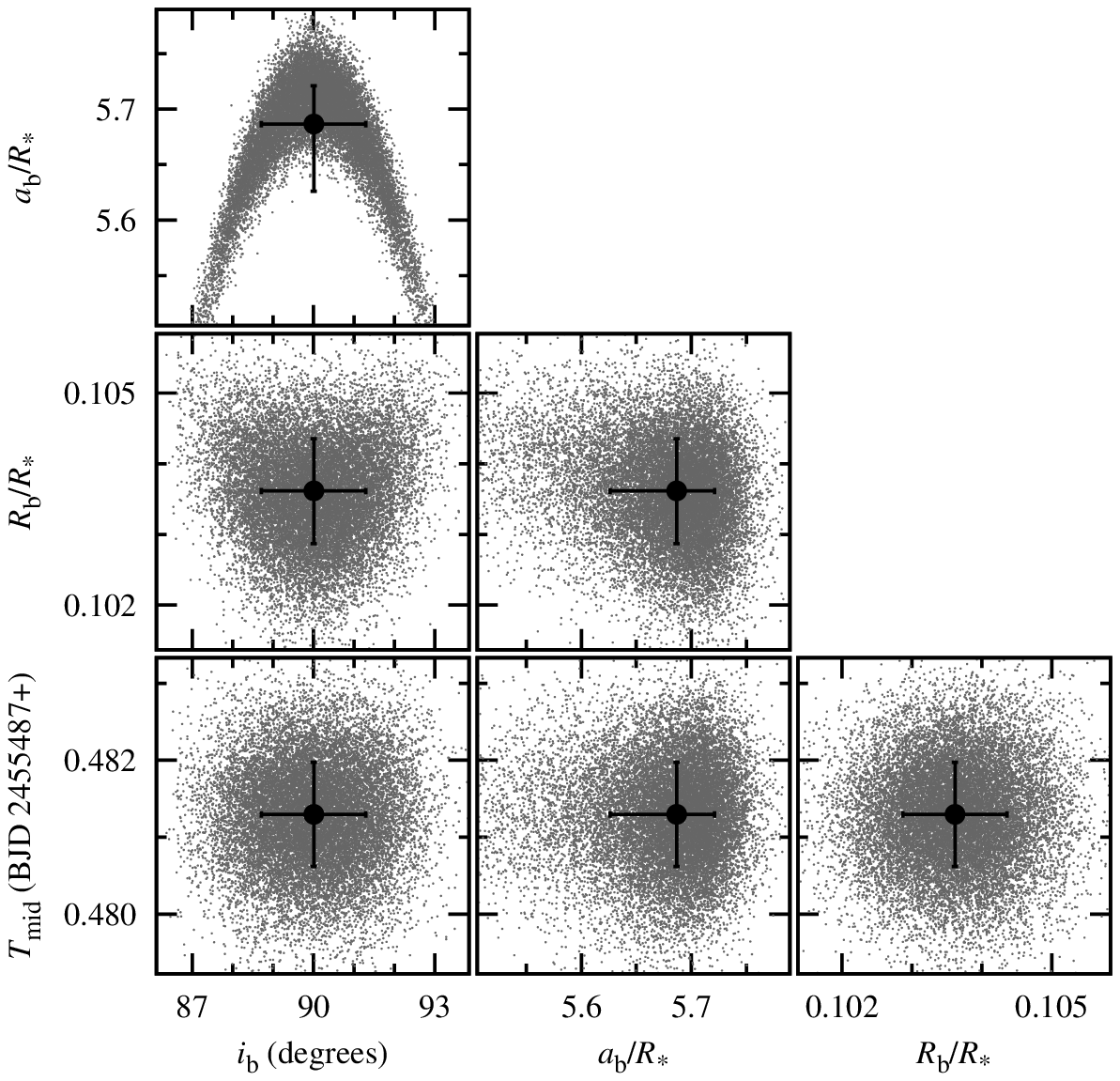}
\end{center}
\FigCap{Relations between fitted light curve parameters from the MCMC iterations, obtained for the high-quality light curve acquired with the 1.2-m telescope in Trebur on 2010 Oct 17. For clarity, every  50th point is plotted. The best fitting parameters are marked with dots.} 
\end{figure}

\begin{figure}[htb]
\begin{center}
\includegraphics[width=1.0\textwidth]{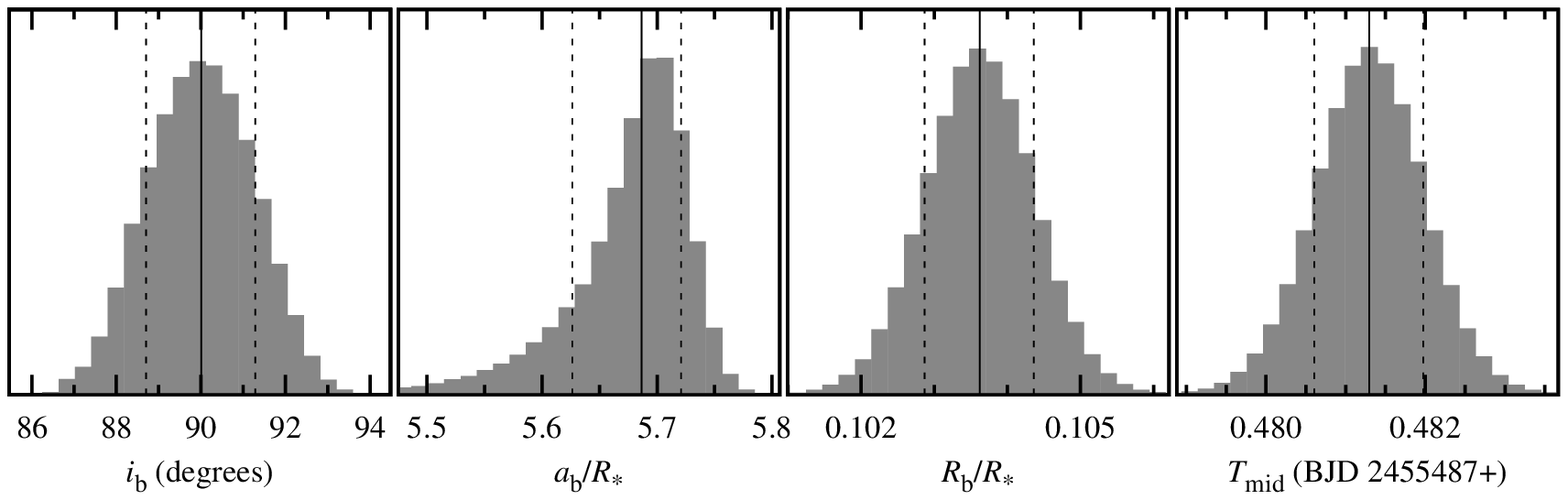}
\end{center}
\FigCap{Histograms of the 10 chains with $10^6$ steps for MCMC distributions of fitted parameters. The distribution of mid-transit times ($T_{\rm{mid}}$) is given for the light curve observed with the 1.2-m telescope in Trebur on 2010 Oct 17. The vertical continuous lines mark the best-fit values, and the dashed lines mark 1-$\sigma$ uncertainties.}
\end{figure}


\subsection{Transit timing}\label{Sect.Timing}

New mid-transit times together with redetermined ones from the literature data are listed in Table~4. They were used to refine the transit ephemeris. A linear fit, which uses individual timing errors as weights, results in  the orbital period
\begin{equation}
 P_{\rm{b}}=2.5199454\pm0.0000005~\rm{d}
\end{equation}
and the mid-transit time at cycle zero
\begin{equation}
  T_{0}=2453912.51510\pm0.00027~\rm{BJD_{TDB}}\, . \;
\end{equation}
We used cycle numbering with epoch 0 starting from the ephemeris given by Collier Cameron \etal (2007). The reduced $\chi^2$ was found to be $0.53$ with $rms_{\rm{ttv}}$ equal to $0.0012$~d. This result shows that transit timing is consistent with the linear ephemeris. The diagram showing differences between observed mid-transit times and those predicted by the linear ephemeris (so called observed minus calculated or O--C diagram) is plotted in Fig.~4.

\MakeTable{lcclc}{12.5cm}{Transit times for WASP-1~b. The value of O--C is the difference between observed and predicted mid-transit times. The source of photometric data is given in the last column.}
{\hline
Date UT & Epoch & $T_{\rm{mid}}$ (BJD$_{\rm{TDB}}$) & ~~O--C (d) & Data \\ 
\hline
2006 Sep 06 &  37 & $2454005.75293^{+0.00049}_{-0.00047}$ & $-0.00015$ & Charbonneau \etal (2007)\\
2006 Oct 04 &  40 & $2454013.31331^{+0.00082}_{-0.00081}$ & $+0.00040$ &  Shporer \etal (2008)\\
2006 Oct 09 &  42 & $2454018.3539^{+0.0010}_{-0.0010}$ & $+0.0011$ &  Shporer \etal (2008)\\
2007 Sep 10 &  361 & $2454353.5026^{+0.0028}_{-0.0024}$ & $-0.0029$ & This work\\
2008 Sep 27 &  327 & $2454736.5368^{+0.0010}_{-0.0010}$ & $-0.0005$ & This work\\
2008 Dec 21 &  361 & $2454822.2148^{+0.0023}_{-0.0022}$ & $-0.0006$ & This work\\
2009 Aug 02 & 450 & $2455046.4895^{+0.0013}_{-0.0013}$ & $-0.0010$ & This work\\
2009 Sep 17 & 468 & $2455091.85005^{+0.00080}_{-0.00081}$ & $+0.00052$ &  Albrecht \etal (2011)\\
2009 Sep 29 & 473 & $2455104.4474^{+0.0020}_{-0.0019}$ & $-0.0019$ & This work\\
2009 Nov 21 & 494 & $2455157.3694^{+0.0012}_{-0.0013}$ & $+0.0013$ & This work\\
2010 Aug 26 & 604 & $2455434.56161^{+0.00052}_{-0.00051}$ & $-0.00049$ & ETD\\
2010 Sep 30 & 618 & $2455469.8411^{+0.0012}_{-0.0012}$ & $-0.0002$ &  Albrecht \etal (2011)\\
2010 Oct 17 & 625 & $2455487.48130^{+0.00068}_{-0.00069}$ & $+0.00034$ & This work\\
2013 Sep 07 & 1044 & $2456543.3381^{+0.0034}_{-0.0030}$ & $+0.0001$ & This work\\
2013 Oct 03 & 1054 & $2456568.5378^{+0.0007}_{-0.0007}$ & $+0.0003$ & This work\\
2013 Oct 08 & 1056 & $2456573.5747^{+0.0035}_{-0.0036}$ & $-0.0027$ & This work\\
2013 Nov 09 & 1069 & $2456606.3371^{+0.0017}_{-0.0016}$ & $+0.0005$ & This work\\
\hline
}

\begin{figure}[htb]
\begin{center}
\includegraphics[width=1.0\textwidth]{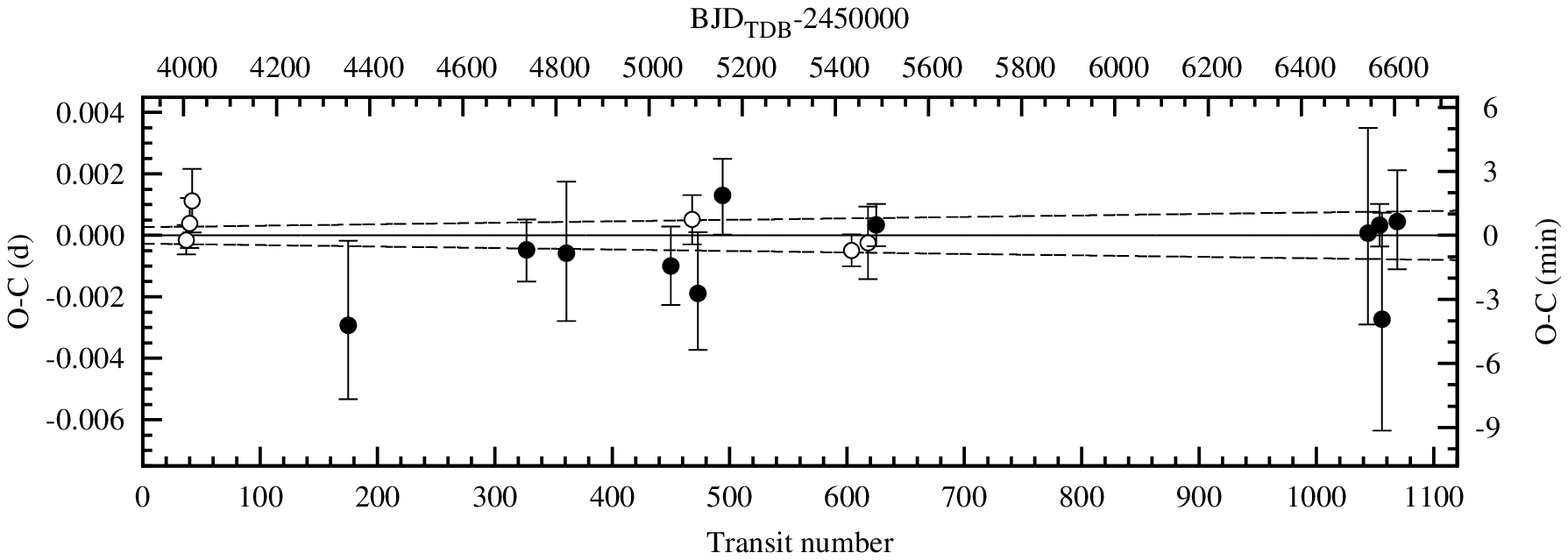}
\end{center}
\FigCap{Residuals from the linear ephemeris for transit timing of WASP-1~b. The open symbols denote reanalysed data from previous studies. The filled dots mark new mid-transit times presented in this paper. The continuous line plots the reference zero level, and the dashed ones show the propagation of 1-$\sigma$ uncertainties of the refined linear ephemeris.}
\end{figure}


\subsection{System parameters}\label{Sect.SysParameters}

\begin{figure}[htb]
\begin{center}
\includegraphics[width=0.6\textwidth]{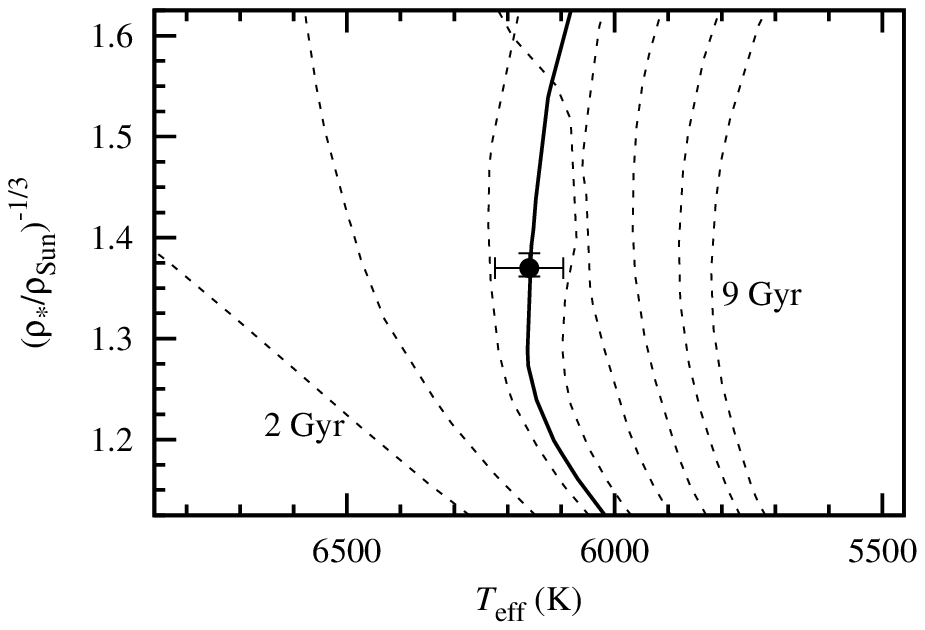}
\end{center}
\FigCap{Position of WASP-1 (the central dot) in the modified Hertzsprung-Russel diagram. The {\sc PARSEC} isochrones of the ages between 2 and 9 Gyr with a step of 1  Gyr are sketched with dashed lines. The best-fitting isochrone is drawn with a continuous line. For clarity, only isochrones interpolated for iron abundance of $\rm{[Fe/H]}=+0.14$ are plotted.}
\end{figure}

The effective temperature of WASP-1 $T_{\rm{eff}}=6160\pm64$~K and the metallicity represented by the iron abundance $\rm{[Fe/H]}=+0.14\pm0.07$ dex, which was taken from Torres \etal (2012), together with the mean stellar density $\rho_{*}$ derived from modeling transit light curves (Sect.~3.1) were used to redetermine the stellar mass $M_{*}$, luminosity $L_{*}$, and age. For planetary systems with transiting planets, the stellar density is determined independently from theoretical stellar models, and can be used instead of the surface gravity ($\log g_{*}$  in cgs units) which is usually determined from spectroscopy with greater uncertainties (see Torres \etal 2012). We employed {\sc PARSEC} isochrones in version 1.1 (Bressan \etal 2012) that were bilinearly interpolated to estimate the parameters and their errors. Figure~5 illustrates the location of WASP-1 in a modified Hertzsprung-Russell diagram together with the best-fitting isochrone. 

The refined value of the stellar mass allowed us to reanalyse RVs from Collier Cameron \etal (2007) and Simpson \etal (2011). The RV measurements come from the SOPHIE spectrograph mounted on the 1.9-m telescope at the Observatoire de Haute Provence (France) and the FIES spectrograph mounted on the 2.6-m Nordic Optical Telescope at the Observatorio del Roque de los Muchachos on La Palma (Spain). The data points that were obtained during transits ($\pm120$ min from the expected mid-transit time, based on the transit ephemeris refined in Sect.~3.2) were removed from the sample to exclude deviations caused by the Rossiter--McLaughlin effect. We used the Systemic Console software (Meschiari \etal 2009) to redetermine the RV semi-amplitude $k_{\rm{b}}$, the minimal planetary mass $M_{\rm{b}} \sin i_{\rm{b}}$, and the semi-major axis, $a_{\rm{b}}$. The mid-transit times from Table~4 were included in the RV model to tighter constrain the orbital mean anomaly at the initial epoch. The orbital period was a free parameter to verify the value obtained from transit timing alone, and to take into account its uncertainty when computing the uncertainties in other parameters. To account for differences in the calibration of system velocities, the RV offsets were allowed to vary for both data sets. In the final iteration, a perfectly circular orbit was assumed. In test runs, when the orbital eccentricity was a free parameter, it quickly converged to zero, leaving no statistically significant hint of an eccentric orbit solution. The Nelder-Mead minimization procedure was used to find the best-fitting Keplerian orbit. The MCMC algorithm was used to determine the parameter uncertainties. The MCMC chain was $10^6$ steps long, and the first 10\% configurations were discarded. The scale parameters were set empirically in a series of attempts to get the acceptance rate of the MCMC procedure close to the optimal value of $0.25$. For each parameter, the standard deviation was taken as the final uncertainty estimate. The scatter of the RVs, characterised by $rms_{\rm{rv}}$, was found to be equal to $16.7$~m~s$^{-1}$.

The determined quantities were combined to calculate the physical parameters of the planet and host star, such as the planetary mass $M_{\rm{b}}$, the radius $R_{\rm{b}}$, the mean density $\rho_{\rm{b}}$, the surface gravitational acceleration $g_{\rm{b}}$, the stellar radius $R_{*}$, and the surface gravity $\log g_{*}$. The value of $g_{\rm{b}}$ was calculated with Eq.\ (7) of Southworth \etal (2008), adapted for a circular orbit 
\begin{equation}
     g_{\rm{b}} = \frac{2\pi}{P_{\rm{b}}}\frac{a_{\rm{b}}^2 k_{\rm{b}}}{R_{\rm{b}}^{2} \sin i_{\rm{b}}}\, . \; 
\end{equation}
The refined system parameters are collected in Table~3.


\subsection{Constraints on the mass of an additional planet}

Transit timing and RV observations allow us to place constraints on the properties of any hypothetical additional planet in the system. We used the {\sc Mercury 6} package (Chambers 1999) with the Bulirsch--Stoer integrator to generate a set of synthetic O--C diagrams for WASP-1~b in the presence of a fictitious perturbing planet. The mass of this body was set at 0.5, 1, 5, 10, 50, 100, and 500 $M_{\rm{Earth}}$ (Earth masses), and the initial semi-major axis varied between 0.0083 and 0.115 AU with a step of $10^{-6}$ AU. The initial orbital longitude of WASP-1~b was set at a value calculated for cycle zero, and the initial longitude of the fictitious planet was shifted by $180^{\circ}$. The system was assumed to be coplanar, with both orbits initially circular.  The integration time covered 2700 days, i.e., the time span of the transit observations. The value of $rms$ of the residuals from a linear ephemeris was calculated for each set of simulated observations. Then, for each orbital distance, we determined the range of planet masses in which the calculated $rms_{\rm{ttv}}$ of $0.0012$~d fell. An upper mass of the fictitious planet at the detection limit was found by linear interpolation for masses below 500 $M_{\rm{Earth}}$. If $rms_{\rm{ttv}}$ was found to be generated by a more massive body, the limiting mass was extrapolated using a linear trend as fitted to 100 and 500 $M_{\rm{Earth}}$. In addition, the value of $rms_{\rm{rv}}$ was used to calculate the RV mass limit as a function of the semi-major axis of the fictitious planet. Finally, both criteria were combined together. The results are illustrated in Fig.~6. 

\begin{figure}[htb]
\begin{center}
\includegraphics[width=0.7\textwidth]{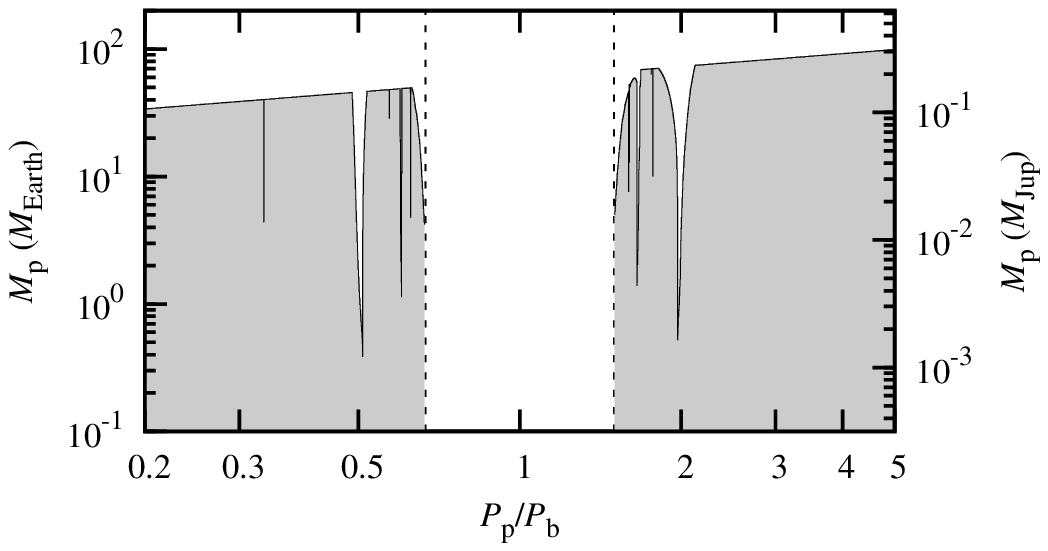}
\end{center}
\FigCap{Upper mass limit for a fictitious additional planet in the WASP-1 system, based on timing and RV data sets, as a function of the orbital period of that planet, $P_{\rm{p}}$. The greyed space of parameters area show configurations that are below our detection threshold. Planetary systems between the inner 2:3 and outer 3:2 orbital period commensurabilities (marked with vertical dashed lines) were found to be highly unstable in most cases.}
\end{figure}

Most of orbits located between the inner 2:3 and outer 3:2 orbital period commensurabilities, were found to be highly unstable and planetary close encounters or planet ejections occurred during the relatively short time of integration. This space of instability was omitted in further analysis. The RV dataset puts tighter constraints in most orbital configurations and allows us to exclude a fictitious planet down to $\sim$30 $M_{\rm{Earth}}$ on $\sim$12-hour orbits. The timing technique allows us to probe the Earth-mass regime close to MMRs. We can exclude an additional planet of the mass of 4.4, 0.4, and 1.1 $M_{\rm{Earth}}$ in inner 1:3, 1:2, and 3:5 MMRs, respectively. We also find no sign of a perturber of the mass greater than 1.4 and 0.5 $M_{\rm{Earth}}$ in outer 5:3 and 2:1 MMRs, respectively.


\section{Conclusions}

Our simultaneous observations of transits with two telescopes at different locations reduced mid-transit time uncertainties by 7--20\%, depending on the quality of individual datasets. Our pilot observations without any filter, \ie in white light, if performed in photometric conditions, result in higher data quality, and hence bring more precise mid-transit times.

The simultaneous analysis of our 13 new transit light curves and additional 6 from the literature results in redetermining system parameters which in most cases agree with the literature ones within $1\sigma$. Despite the conservative approach, we derived more precise values than those ones published before.   

The homogeneous analysis of transit light curves shows that transit times of the WASP-1~b planet follow the linear ephemeris, leaving no space for any sign of timing variation. Thus, we can only put observational constraints for a fictitious second planet in the system. The transiting planet has no detectable close planetary companion and this finding agrees with the analysis of a sample of hot Jupiter candidates observed with the {\it Kepler} space telescope.


\Acknow{We thank Mohammad Moualla for trying to obtain additional data. GM and DP acknowledge the financial support from the Polish Ministry of Science and Higher Education through the Iuventus Plus grant IP2011 031971. 
JN and MV would like to thank the project
VEGA 2/0143/14 and the project APVV-0158-11. 
CM thanks the German national science foundation Deutsche Forschungsgemeinschaft
(DFG) for support through the grant SCHR665/7-1.
SR acknowledges the financial support from DFG in the grant NE 515/33-1. 
We acknowledge financial support from the Thuringian government (B 515-07010) for the STK CCD camera used in this project. 
Part of this paper is the result of the exchange and joint research project {\em Spectral and photometric studies of variable stars} between Polish and Bulgarian Academies of Sciences. This research has made use of the Exoplanet Transit Database, maintained by Variable Star Section of Czech Astronomical Society.}

\end{document}